\documentclass[aps,pra,preprint,showpacs,superscriptaddress,groupedaddress,notitlepage]{revtex4-1}

\usepackage{graphicx}

\usepackage{dcolumn}   
\usepackage{bm}        

\usepackage{subfig}
\usepackage{amsmath}	
\usepackage{amssymb}    

\newcommand{\be}{\begin{eqnarray}}
 \newcommand{\ee}{\end{eqnarray}}

\begin{document}



\title{Landau-Zener-St\"uckelberg interferometry in pair production from counterpropagating lasers}

\author{Fran\c{c}ois Fillion-Gourdeau}
\email{filliong@CRM.UMontreal.ca}
\affiliation{Centre de Recherches Math\'{e}matiques, Universit\'{e} de Montr\'{e}al, Montr\'{e}al, Canada, H3T 1J4}
\altaffiliation[Also at ]{School of Mathematics and Statistics, Carleton University, Ottawa, Canada, K1S 5B6; and \\Fields Institute, University of Toronto, Toronto, Canada, M5T 3J1}

\author{Emmanuel Lorin}
\email{elorin@math.carleton.ca}
\affiliation{School of Mathematics and Statistics, Carleton University, Ottawa, Canada, K1S 5B6}
\altaffiliation[Also at ]{Centre de Recherches Math\'{e}matiques, Universit\'{e} de Montr\'{e}al, Montr\'{e}al, Canada, H3T 1J4}

\author{Andr\'{e} D. Bandrauk}
\email{Andre.Dieter.Bandrauk@USherbrooke.ca}
\affiliation{Laboratoire de chimie th\'{e}orique, Facult\'{e} des Sciences, Universit\'{e} de Sherbrooke, Sherbrooke, Canada, J1K 2R1}
\altaffiliation[Also at ]{Centre de Recherches Math\'{e}matiques, Universit\'{e} de Montr\'{e}al, Montr\'{e}al, Canada, H3T 1J4}

\date{\today}

\begin{abstract}
The rate of electron-positron pair production in linearly polarized counter-propagating lasers is evaluated from a recently discovered solution of the time-dependent Dirac equation. The latter is solved in momentum space where it is formally equivalent to the Schr\"odinger equation describing a strongly driven two-level system. The solution is found from a simple transformation of the Dirac equation and is given in compact form in terms of the doubly-confluent Heun's function. By using the analogy with the two-level system, it is shown that for high-intensity lasers, pair production occurs through periodic non-adiabatic transitions when the adiabatic energy gap is minimal. These transitions give rise to an intricate interference pattern in the pair spectrum, reminiscent of the Landau-Zener-St\"uckelberg phenomenon in molecular physics: the accumulated phase result in constructive or destructive interference. The adiabatic-impulse model is used to study this phenomenon and shows an excellent agreement with the exact result. 
\end{abstract}


\maketitle




\section{Introduction}

The production of electron-positron pairs from classical external fields has a long history, starting with the seminal work of Schwinger \cite{PhysRev.82.664,Itzykson:1980rh}, where antimatter production from a constant electric field was considered. On the theoretical side, Schwinger's mechanism is relatively well understood and is usually interpreted as the decay of the vacuum into a particle-antiparticle pairs (in the Dirac sea picture, this is seen a a tunneling from the negative to the positive energy sea). However, an experimental validation of this phenomenon is still out of reach:  the intensity of a laser electric field required to produce an observable amount of pairs is on the order of $10^{29}$ W/cm$^{2}$ \cite{Salamin200641}, which is still unattainable experimentally. This occurs because Schwinger's result states that the probability to produce a pair (per unit volume and time) results from a tunnelling process and is given by \cite{PhysRev.82.664,Itzykson:1980rh}
\begin{eqnarray}
P_{S} \sim e^{- \frac{\pi m^{2} c^{3}}{|e|E}},
\end{eqnarray}
where $m$ is the electron mass, $c$ the speed of light, $|e|$ the absolute value of the electron charge and $E$ the electric field strength. Thus, an appreciable amount of pairs can be produced only if $E \sim E_{S} \equiv \frac{m^{2} c^{3}}{e} \approx 10^{18}$ V/m, which is much larger than $E_{1s} = 5 \times 10^{11}$ V/m, the typical electric field in the ground state of an atom (1s orbital).

 In the last few decades, laser technologies have made a giant leap forward such that electric field of unprecedented intensity level can be attained (on the order of $10^{22}$ W/cm$^{2}$ and higher \cite{RevModPhys.78.309}). At these intensities, relativistic effects start to be important and thus, the spontaneous creation of electron-positron pairs from laser fields becomes more plausible. This has triggered many theoretical studies recently where pair production from generalizations or variations of Schwinger's process were considered. Thus, different field configurations to produce pairs have been studied such as counterpropagating lasers \cite{PhysRevD.2.1191,PROP:PROP19770250111,Nikishov1973,PhysRevE.66.016502}, counterpropagating lasers with space dependence \cite{PhysRevLett.102.080402}, laser field with heavy nuclei \cite{PhysRevA.67.063407,PhysRevLett.100.010403,PhysRevLett.91.223601} and the combination of rapidly and slowly varying fields \cite{PhysRevA.85.033408,PhysRevD.80.111301}. The effect of the temporal laser pulse shape has also been investigated \cite{PhysRevD.70.053013,PhysRevLett.104.250402,PhysRevD.83.065028}.  

Although the machinery for computing the number of pairs produced in the strong external field approximation is well-known \cite{Nikishov1964,Greiner:1985}, its evaluation is still a challenging task because it is related to a solution of the Dirac equation, which is notoriously hard to solve. For this reason, most of the analytical studies have focused on simple systems. For instance, the pair production from a time-varying homogeneous electric field was treated in \cite{PhysRevD.2.1191,PROP:PROP19770250111,Nikishov1973}, using different schemes of approximation. The main goal of this article is to revisit this problem from a slightly different perspective: the production of pairs from an electric field representing counterpropagating lasers in the dipole approximation is evaluated by using a solution of the Dirac equation and the adiabatic-impulse model, allowing to study interference effects in the pair production spectrum. 

It was recently argued by using a semi-classical approximation that these interference effects are related to Stoke's phenomenon and are responsible for the peak-valley structure in the pair spectrum \cite{PhysRevD.83.065028,PhysRevLett.108.030401,PhysRevLett.104.250402}. The main justification of this result is that interference occurs between the semi-classical turning points due to the different phases acquired during the time evolution. In this work, this phenomenon is investigated further by using the formal analogy of the Dirac equation describing our system with the Driven Two-Level System (DTLS). It is well-known that in a certain regime (defined later), the time evolution of the DTLS occurs adiabatically, except for some specific times where the system goes through an avoided crossing between the ``dressed'' energy levels. In this case, non-adiabatic transitions take place between the lower and upper energy levels. As the avoided crossing region is passed many times, these transitions interfere and can lead to constructive or destructive interference according to the phase accumulated during the transition and the adiabatic evolution. This phenomenon is named the Landau-Zener-St\"uckelberg Interferometry (LZSI). It is described extensively in \cite{Shevchenko20101}, and references therein, and is important in molecular physics \cite{child2010molecular,doi:10.1080/00268977000101041,nakamura:4031}. In this work, it is shown that the LSZI is also relevant for pair production in high intensity laser field and that pair production in lasers proceeds by periodic non-adiabatic transitions. To make this connection and for simplicity, the 1-D case is considered, which corresponds in 3-D to the production of pairs at zero transverse momentum ($p_{\perp}=0$). Other 1-D models have been considered in the literature \cite{PhysRevD.83.065028,PhysRevLett.108.030401,PhysRevLett.104.250402,PhysRevA.80.062105}.

This article is separated as follows. In Section \ref{sec:pair_form}, the formalism to compute the rate of pairs produced is presented. More precisely, it is shown that the average number of pairs produced is related to coefficients in the solution of the Dirac equation. The latter is solved in Section \ref{sec:sol_dirac} in the background field of linearly polarized counterpropagating lasers. In Section \ref{sec:adia_model}, a theoretical approach to evaluate the wave function approximately in the adiabatic-impulse model is presented. The pair production and numerical results are shown in Section \ref{sec:num_res}, along with an interpretation in terms of nonadiabatic transitions. We take advantage of the analogy with the DTLS and evaluate the number of pair produced with the adiabatic-impulse model, allowing to understand the spectrum in terms of the LZSI. We conclude in Section \ref{sec:conclu}. Throughout this work, we use the metric $g_{\mu \nu} = \mathrm{diag}(1,-1,-1,-1)$. Also, units in which $\hbar = c  = m = 1$ (where $m$ is the electron mass) and $e=\sqrt{\alpha}$ are utilized in most numerical calculations. In this case, the unit length is $l_{\rm u} = \hbar/(mc) \sim 3.86159 \times 10^{-13}$~m (38.6 picometer) while the unit time is $t_{\rm u} = \hbar / (mc^{2}) \sim 1.2880885 \times 10^{-21}$~s (1.288 zeptosecond), as compared to atomic units: $l_{\rm a.u.} = 0.052$ nm and $t_{\rm a.u.} = 2.4 \times 10^{-18}$ s (2.4 attosecond). Note that in these units, the Schwinger field is $|eE_{S}| = 1$.   

\section{Pair production from strong classical fields}
\label{sec:pair_form}

The mathematical description of pair production requires a Quantum Field Theory (QFT) treatment because it involves particle creation and annihilation. The main tool to calculate observable quantities in this framework is a perturbation theory in terms of the coupling constant (Feynman diagrams), which allows to evaluate approximately the value of field correlators. The latter can be linked to physical observables by using reduction formula based on the Lehmann-Symanzik-Zimmermann (LSZ) asymptotic conditions for the field at $t = \mp \infty$. In these limits, the quantized field operator $\hat{\Psi}$ is known and is given by \cite{Fukushima2009184}
\begin{eqnarray}
\hat{\Psi}_{\rm in,out}(x) &=& \lim_{t\rightarrow \mp \infty} \hat{\Psi}(x,t), \\
&=& \int \frac{dp}{(2\pi)} \left[ \frac{\hat{a}_{\rm in,out}(p)}{2E_{p}^{\rm in,out}}u_{\rm in,out}(p)e^{-iE_{p}^{\rm in,out}t+ipx} +  \frac{\hat{b}^{\dagger}_{\rm in,out}(p)}{2E_{-p}^{\rm in,out}}v_{\rm in,out}(p)e^{iE_{-p}^{\rm in,out}t-ipx} \right],
\end{eqnarray}
where $\hat{a}_{\rm in,out}(p),\hat{b}_{\rm in,out}(p)$ are annihilation operators that annihilate the ``in,out'' vacuum as
\begin{eqnarray}
 \hat{a}_{\rm in,out}(p)|0_{\rm in,out} \rangle = \hat{b}_{\rm in,out}(p)|0_{\rm in,out} \rangle = 0 .
 \end{eqnarray}
In these last equations, we have also introduced the asymptotic energies $E_{p}^{\rm in,out}$ and the free positive/negative energy spinors $u_{\rm in,out},v_{\rm in,out}$. The explicit expression of these quantities depends on the form of the electromagnetic potential and the gauge chosen; they will be described precisely below. On the other hand, the final result for the number of pairs produced is gauge invariant.

When a strong external field such as a laser is involved and coupled to the fermionic degrees of freedom, the ``naive'' perturbation series is no longer asymptotically convergent because field insertions, being parametrically of order $1/e$, contribute to leading order. The field insertions have to be resummed to obtain the leading order contribution. For some observables, such as the average number of pair produced from the vacuum $\langle n \rangle$, this resummation can be performed in the Schwinger-Keldysh formalism by using the Lippmann-Schwinger equation \cite{PhysRevC.71.024904,Gelis2006135,Baltz2001395}. The main result of this procedure is a relation between the physical quantity $\langle n \rangle$ and a solution of the ``classical'' Dirac equation in the laser background field.

This relation is the starting point of this work and is given by \cite{PhysRevC.71.024904,Gelis2006135,Baltz2001395}
\begin{eqnarray}
\langle n \rangle &=& \int d\tilde{p}^{+}_{\rm out} d\tilde{q}^{-}_{\rm in} \left| \lim_{t \rightarrow \infty}  e^{iE^{\rm out}_{p}t } u^{\dagger}_{\rm out}(p) {\Psi}_{q}(t,p) \right|^{2},
\label{eq:pair_prod}
\end{eqnarray}
where ${\Psi}_{q}(t,p)$ is the Fourier transform (with respect to space) of $\Psi_{q}(t,\mathbf{x})$, the retarded solution of the 1-D Dirac equation in coordinate space. Here, the subscript refers to the momentum of the initial state: the wave function is subjected to the initial condition 
\begin{eqnarray}
\lim_{t \rightarrow -\infty} {\Psi}_{q}(t,p) = v_{\rm in}(q)e^{iE^{\rm in}_{-q} t} (2\pi) \delta(p+q).
\label{eq:init_fourier}
\end{eqnarray}
It should be noted here that this condition is derived in the resummation procedure \cite{PhysRevC.71.024904,Gelis2006135,Baltz2001395}. Physically, this means that the average number of pair produced is computed by preparing the system in a negative energy state at $t \rightarrow -\infty$, by evolving the wave function in time and projecting it onto a positive energy state at $t \rightarrow \infty$. Although the Dirac equation is solved with an initial state representing a positron, the quantity $\langle n \rangle$ represents the number of pairs produced from the vacuum, without positrons in the initial state. This may seem counter-intuitive but it is related to causality and the fact that $\langle n \rangle$ requires the evaluation of Wightman propagators. This can also be understood in the Feynman-St\"uckelberg interpretation of the positron as an electron evolving backward in time.

In 1-D, the wave function $\Psi$ is a bi-spinor and the Dirac equation obeyed by ${\Psi}_{q}(t,p)$ is then
\begin{equation}
\label{eq:dirac_eq_1D}
 i\partial_{t}{\Psi}(t,p) = \left[  \alpha (cp +A_{x}(t))  + \beta mc^{2}   \right] {\Psi}(t,p), 
\end{equation}
where $m$ is the electron mass. The Dirac equation is expressed in a gauge where the scalar potential $A_{0} = 0$ while the vector potential $A_{x}(t)$ is both time-dependent and space-independent. Throughout this work, we work in a representation where the Dirac matrices are given by Pauli matrices such as $\alpha = \sigma_{z}$ and $\beta = \sigma_{x}$. In such representation, Eq. \eqref{eq:dirac_eq_1D} is supersymmetric \cite{thaller1992dirac} for which semiclassical WKB approximation are exact \cite{Comtet1985159}.       

The covariant measure $d\tilde{p}^{\pm}_{\rm in,out} \equiv \frac{dp}{(2\pi)2E^{\rm in,out}_{\pm p}}$ is defined with respect to the ``asymptotic'' energies 
\begin{eqnarray}
E^{\rm in,out}_{p} \equiv \sqrt{( cp + G^{\rm in,out})^{2} + m^{2}c^{4}}, 
\end{eqnarray}
which are obtained by the on-shell conditions at $t=\pm \infty$. The $G^{\rm in,out}$ are constants related to the gauge potential as $\lim_{t \rightarrow \pm \infty } A_{x}(t) = G^{\rm in,out}$. Thus, although the physical electric field vanishes asymptotically, the potential may have a non-zero value which depends on the gauge chosen.

The spinors $u_{\rm in,out}(p)$ and $v_{\rm in,out}(p)$ are the positive and negative energy solutions of Eq. \eqref{eq:dirac_eq_1D} with $G(t)=G^{\rm in,out}$ and normalized such that $u^{\dagger}_{\rm in,out}(q)u_{\rm in,out}(q) = v_{\rm in,out}^{\dagger}(q)v_{\rm in,out}(q) = 2E_{q}^{\rm in,out}$. Explicitly, they are given by
 \begin{eqnarray}
  u_{\rm in,out}(q) &=& 
 \begin{bmatrix}
\sqrt{E^{\rm in,out}_{q}+(cq+G^{\rm in,out})} \\
  \sqrt{E^{\rm in,out}_{q}-(cq+G^{\rm in,out})}
 \end{bmatrix} , \nonumber \\
 %
 %
  v_{\rm in,out}(q) &=& 
 \begin{bmatrix}
   \sqrt{E^{\rm in,out}_{-q}-(-cq+G^{\rm in,out})} \\
-\sqrt{E^{\rm in,out}_{-q}+(-cq+G^{\rm in,out})}  
 \end{bmatrix} .
 \end{eqnarray}
As usual, the index $\rm in,out$ represents the limits at $t \rightarrow \mp \infty$.

So far, we have defined a general expression for $\langle n \rangle$. We now specializes this to the case of an external laser field. More precisely, the electric field considered is given by
\begin{eqnarray}
 E(t)=E\sin(\omega t) = -\frac{\partial A_{x}(t)}{\partial t},
 \end{eqnarray}
where $E$ is the field strength and $\omega$ is the laser frequency. It represents the field from linearly polarized counterpropagating lasers where the space variations are neglected. In other words, we consider pair production in the neighborhood of the standing wave anti-nodes, where the electric field reaches its maximum value. This field is applied during a time interval $t \in [0,T]$ where $T$ is the final time. By working in a gauge where the scalar potential is zero ($A_{0}=0$), the vector potential is given by
\begin{eqnarray}
\label{eq:potential}
 A_{x}(t) = 
\begin{cases}
 G(0)\equiv G^{\rm in}, & t \in (-\infty,0] \\
 G(t), & t \in [0,T] \\
G(T)\equiv G^{\rm out}, & t \in [T,\infty) 
\end{cases},
\end{eqnarray}
where $G(t) \equiv -\int^t E(t')dt' = \frac{F}{\omega}\cos(\omega t)$, with $F$ the normalized field strength (normalized as $F\equiv |e|cE$).

The solution of the Dirac equation in the potential considered can then be written as
\begin{eqnarray}
\label{eq:wf_total}
 {\Psi}_{q}(t,p) = 
\begin{cases}
 v_{\rm in}(q)e^{iE^{\rm in}_{-q} t} (2\pi) \delta(p+q), & t \in (-\infty,0] \\
 {\psi}(t,p), & t \in [0,T] \\
A u_{\rm out}(p)e^{-iE^{\rm out}_{p} t} 
+ B v_{\rm out}(-p)e^{iE^{\rm out}_{p} t}, & t \in [T,\infty) 
\end{cases},
\end{eqnarray}
where ${\psi}(t,p)$ is a solution of the Dirac equation with the laser vector potential, $A,B$ are integration constants that need to be determined from the data at $t=T$ (thus, their value depend on $T$ and $p$) and which allows to have a linear combination of negative and positive energy free solutions. Thus, initially, the wave function is given by $v_{\rm in}$ which represents a positron. Substituting the last equation in Eq. \eqref{eq:pair_prod}, taking the limit, using the fact that $u^{\dagger}(p)v(-p)=0$ and $u^{\dagger}(p)u(p)=2E_{p}$, and integrating on the positron momentum (using the delta function of the initial state, related to translation invariance), we get 
 \begin{eqnarray}
\langle n (T)\rangle &=& \frac{V}{2\pi}\int dp   \frac{E^{\rm out}_{p}}{E^{\rm in}_{p}}\left| A(T,p) \right|^{2}.
\label{eq:pair_prod_1D}
\end{eqnarray}
where $V = \delta(0)$ is the infinite volume. As usual, this diverging quantity is treated by redefining $\langle n \rangle$ as the number of pair produced per unit volume. This is the convention used in the rest of this work. 

By requiring the continuity of the solution at $t=T$, we get the following conditions:
\begin{eqnarray}
{\psi}(T,p) = A(T,p) u_{\rm out}(p)e^{-iE^{\rm out}_{p} T} + B(t,p) v_{\rm out}(-p)e^{iE^{\rm out}_{p} T}.
\end{eqnarray}
This can be used to compute the constant $A$ which is directly related to pair production via Eq. \eqref{eq:pair_prod_1D}. It is a straightforward calculation to obtain
\begin{eqnarray}
A(T,p) &=& \left[ \frac{u_{\mathrm{out},1}(p){\psi}_{1}(T,p) + u_{\mathrm{out},2}(p){\psi}_{2}(T,p)}{2E^{\rm out}_{p}} \right] e^{iE^{\rm out}_{p}T} .
\end{eqnarray} 
Thus, we have all the ingredients to calculate the pair production from the vacuum in a counterpropagating laser field. In the following, the coefficient $A$ will be evaluated numerically to obtain the average number of pair produced. The calculation starts by obtaining an exact solution of the Dirac equation with the time-dependent background field: the wave function $\psi(T,p)$ has to be determined.

\section{Solution of the Dirac equation}
\label{sec:sol_dirac}

Substituting the potential defined in Eq. \eqref{eq:potential} for $t \in [0,T]$ in Eq. \eqref{eq:dirac_eq_1D} yields the following Dirac equation:
\begin{equation}
\label{eq:dirac_eq_pot}
 i\frac{d}{dt} \psi(t,p) = \left[  \sigma_{z} \left(cp + \frac{F}{\omega} \cos(\omega t) \right)  + \sigma_{x} mc^{2}   \right] {\psi}(t,p). 
\end{equation}
It should be noted here that this is identical to the equation describing the strongly periodically driven two-level system \cite{Grifoni1998229,Shevchenko20101} and is also in the supersymmetric form of the Dirac equation \cite{thaller1992dirac,Comtet1985159}. The formal analogy between the two systems is recovered by letting $mc^{2} \rightarrow -\Delta/2$, $cp \rightarrow \epsilon_{0} $ and $ \frac{F}{\omega}  \rightarrow  A $ (in the notation of \cite{Shevchenko20101}). This means that for each momentum $p$ corresponds a different two-level system.


Using the explicit expression for Dirac/Pauli matrices, the last equation can be written componentwise as
\begin{eqnarray}
\label{eq:dirac_comp}
 \left[ i\frac{d}{dt} \mp  \left(cp + \frac{F}{\omega} \cos(\omega t) \right)  \right] {\psi}_{1,2}(t,p) =  mc^{2}    {\psi}_{2,1}(t,p). 
\end{eqnarray}
These two equations can be decoupled easily to get the following system of differential equations:
\begin{eqnarray}
\label{eq:dirac_eq_sec}
\biggl[ \frac{d^{2}}{dt^{2}} - i F\sin(\omega t)  +  \left(cp+\frac{F}{\omega}\cos(\omega t) \right)^{2}  
 + m^{2}c^{4} \biggr]{\psi}_{1}(t,p)=0,\\
\label{eq:dirac_eq_sec2}
{\psi}_{2}(t,p) = \frac{1}{mc^{2}} \left[ i\partial_{t} -  \left(cp + \frac{F}{\omega} \cos(\omega t) \right)  \right] {\psi}_{1}(t,p) .
\end{eqnarray}
Eq. \eqref{eq:dirac_eq_sec} is a second order Hill's differential equation. It is to be noted that Eq. \eqref{eq:dirac_eq_sec} contains an explicit imaginary part as in optical potential problems, corresponding to absorption out of the wave function $\psi_{1}$ into the state $\psi_{2}$, i.e. a non-adiabatic transition \cite{McCann1990509}. As demonstrated in \cite{PhysRevA.82.032117}, the last equation can be solved analytically in terms of Heun's function (similar equations were also treated in \cite{1751-8121-44-47-475304,0305-4470-35-12-312}), allowing to evaluate the first component of the wave function $\psi_{1}$. The second component can then be found by substituting $\psi_{1}$ into Eq. \eqref{eq:dirac_eq_sec2}. This way of calculating the solution insures that the general solutions of Eqs. \eqref{eq:dirac_eq_sec} and \eqref{eq:dirac_eq_sec2} are also solutions of the first order system of equations given in Eq. \eqref{eq:dirac_comp}, at all times. 

\begin{figure}
\includegraphics[width=0.8\textwidth]{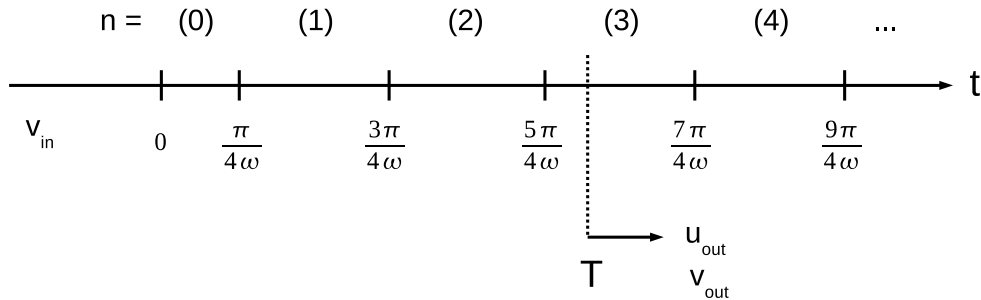}
\caption{The domain $\mathbb{R}^{+}$ for $t$ is separated into subdomains parametrized by $n \in \mathbb{Z}^{+}$ such that $t \in \left[ \frac{(2n-1)\pi}{4\omega}, \frac{(2n+1)\pi}{4\omega}\right]$. At the final time $t=T$, the wave function is matched to the free solution $u_{\rm out},v_{\rm out}$. }
\label{fig:domain}
\end{figure}

To solve Eq. \eqref{eq:dirac_eq_sec}, we follow a strategy similar to \cite{PhysRevA.82.032117}: the time domain $\mathbb{R}^{+}$ is separated into subdomains of length $\pi/2\omega$, parametrized by a positive integer $n$ (see Fig. \ref{fig:domain}). Then, Eq. \eqref{eq:dirac_eq_sec} is solved for $t \in \Delta t_{n} = \left[ \frac{(2n-1)\pi}{4\omega}, \frac{(2n+1)\pi}{4\omega}\right]$; the general solution at all times is determined by matching solutions at points $t_{n} = \frac{(2n-1)\pi}{4\omega}$ for each interval, using continuity conditions.   The rationale behind this procedure is related to the convergence radius of Heun's functions, as will be clarified later.

The following change of variable is then applied to Eq. \eqref{eq:dirac_eq_sec} for $t \in \Delta t_{n}$:
\begin{eqnarray}
\label{eq:trans}
 z = -i \tan \left( \frac{\omega t}{2} - \frac{n\pi}{4}  \right).
\end{eqnarray}
This transformation maps the time interval $\Delta t_{n}$ onto a new variable $iz \in [-\tan (\pi/8),\tan (\pi/8)]$. The inverse transformation is given by
\begin{eqnarray}
 t = \frac{2}{\omega} \arctan (iz) + \frac{n\pi}{2\omega},
\end{eqnarray}
where the $\arctan$ function should be evaluated on its principal value. Using this prescription and the transformation in Eq. \eqref{eq:trans} on the domain $\Delta t_{n}$, the change of variable is bijective and well-defined. A differential equation in terms of the variable $z$ and the model parameters is obtained (not shown here for simplicity). The latter is solved by seeking solutions of the form:
\begin{eqnarray}
 \psi_{1}(z) =\exp \left[- e^{i\frac{n\pi}{2}}  \frac{2F}{\omega^{2}} \frac{z}{z^{2} - 1}\right]  H^{(n)}_{a}(z),
\end{eqnarray}
where $H^{(n)}_{a}(z)$ is the first linearly independent solution (the second linearly independent solution will be denoted by $H^{(n)}_{b}(z)$). These two transformations convert Eq. \eqref{eq:dirac_eq_sec} into a double confluent Heun's equation \cite{ronveaux1995heun}:
\begin{eqnarray}
\label{eq:heun_diff}
 \biggl[ \frac{d^{2}}{dz^{2}} - \frac{-2z^{5} + 4z^{3} + \alpha^{(n)} z^{4} - 2z -\alpha^{(n)}}{(z-1)^{3}(z+1)^{3}}\frac{d}{dz} 
+ \frac{z^{2}\beta^{(n)} +(\gamma^{(n)} + 2 \alpha^{(n)})z + \delta^{(n)} }{(z-1)^{3}(z+1)^{3}} \biggr] H^{(n)}_{a}(z) = 0, \nonumber \\
\end{eqnarray}
where $\alpha^{(n)},\beta^{(n)},\gamma^{(n)},\delta^{(n)}$ are parameters which value depends on the interval considered. They are given explicitly by
\begin{eqnarray}
 \alpha^{(n)} &=&   -e^{i\frac{n\pi}{2}}{\frac {4F}{\omega^{2}}}, \\
\beta^{(n)} &=& {\frac {8Fcp\cos \left( \frac{n\pi}{2}  \right) 
\omega-4iF\sin \left( \frac{n\pi}{2}  \right) \omega^{2}-4\omega^{2}({c}^{2}{
p}^{2}+ {m}^{2}{c}^{4})+2{F}^{2} [\left( -1 \right) ^{
n}-1]}{\omega^{4}}},\\
\gamma^{(n)} &=& {\frac {-16i
Fcp\sin \left( \frac{n\pi}{2}  \right) +8F\cos \left( \frac{n\pi}{2} 
 \right) \omega}{\omega^{3}}}, \\
\delta^{(n)} &=& {\frac{8Fcp\cos \left( \frac{n\pi}{2}  \right) \omega-4iF\sin \left( \frac{n\pi}{2}  \right) \omega^{2}+4\omega^{2}({c}^{2}{p}^{2}+{m}^{2}{c}
^{4})-2{F}^{2} [\left( -1 \right) ^{n} -1]}{\omega^{4}}}.
\end{eqnarray}
Eq. \eqref{eq:heun_diff} has singularities at $z=\pm 1$ while the point $z=0$ is regular. This allows to obtain a power series solution around $z=0$ with a radius of convergence defined by the condition $|z|<1$. Note here that this condition is always fulfilled for all intervals $\Delta t_{n}$ because $iz \in [-\tan (\pi/8),\tan (\pi/8)] \approx [-0.414,0.414]$.  This power series is well-known and the solution is given by \cite{ronveaux1995heun}
\begin{eqnarray}
 H^{(n)}_{a}(z) = H_{D}(\alpha^{(n)},\beta^{(n)},\gamma^{(n)},\delta^{(n)},z),
\end{eqnarray}
where $H_{D}$ is the doubly confluent Heun's function. The partition of the domain has been chosen such that the argument of the Heun function $z$ is always within the radius of convergence, guaranteeing that it is a valid solution and facilitating the numerical evaluation. 

This yields the first particular solution. A second linearly independent solution can be found using a well-known procedure \cite{PhysRevA.82.032117}:
\begin{eqnarray}
 H^{(n)}_{b}(z) = e^{-\alpha^{(n)} \frac{z}{z^{2}-1}}H_{D}(-\alpha^{(n)},\beta^{(n)},\gamma^{(n)},\delta^{(n)},z).
\end{eqnarray}
The final result obtained from this is that the first component of the time-dependent wave function is given by 
\begin{eqnarray}
\label{eq:sol_H1}
{\psi}^{(n)}_{1}(t) &=& A^{(n)} e^{-i e^{\frac{n\pi}{2}} \frac{F}{\omega^{2}} \sin \left(\omega t - \frac{n \pi}{2} \right)} 
H_{D} \left[\alpha^{(n)},\beta^{(n)}, \gamma^{(n)},\delta^{(n)},-i \tan \left( \frac{\omega t}{2} - \frac{n\pi}{4}  \right)  \right] \nonumber \\
&+& B^{(n)} e^{i e^{\frac{n\pi}{2}} \frac{F}{\omega^{2}} \sin \left(\omega t - \frac{n \pi}{2} \right)} 
H_{D} \left[
-\alpha^{(n)},\beta^{(n)}, \gamma^{(n)},\delta^{(n)},
	-i \tan \left( \frac{\omega t}{2} - \frac{n\pi}{4}  \right)  \right] 
\end{eqnarray}
where $A^{(n)},B^{(n)}$ are integration constants that need to be fixed by initial conditions. The second component $\psi_{2}^{(n)}$ is then given by Eq. \eqref{eq:dirac_eq_sec2}.

\begin{figure}
\includegraphics[width=0.9\textwidth]{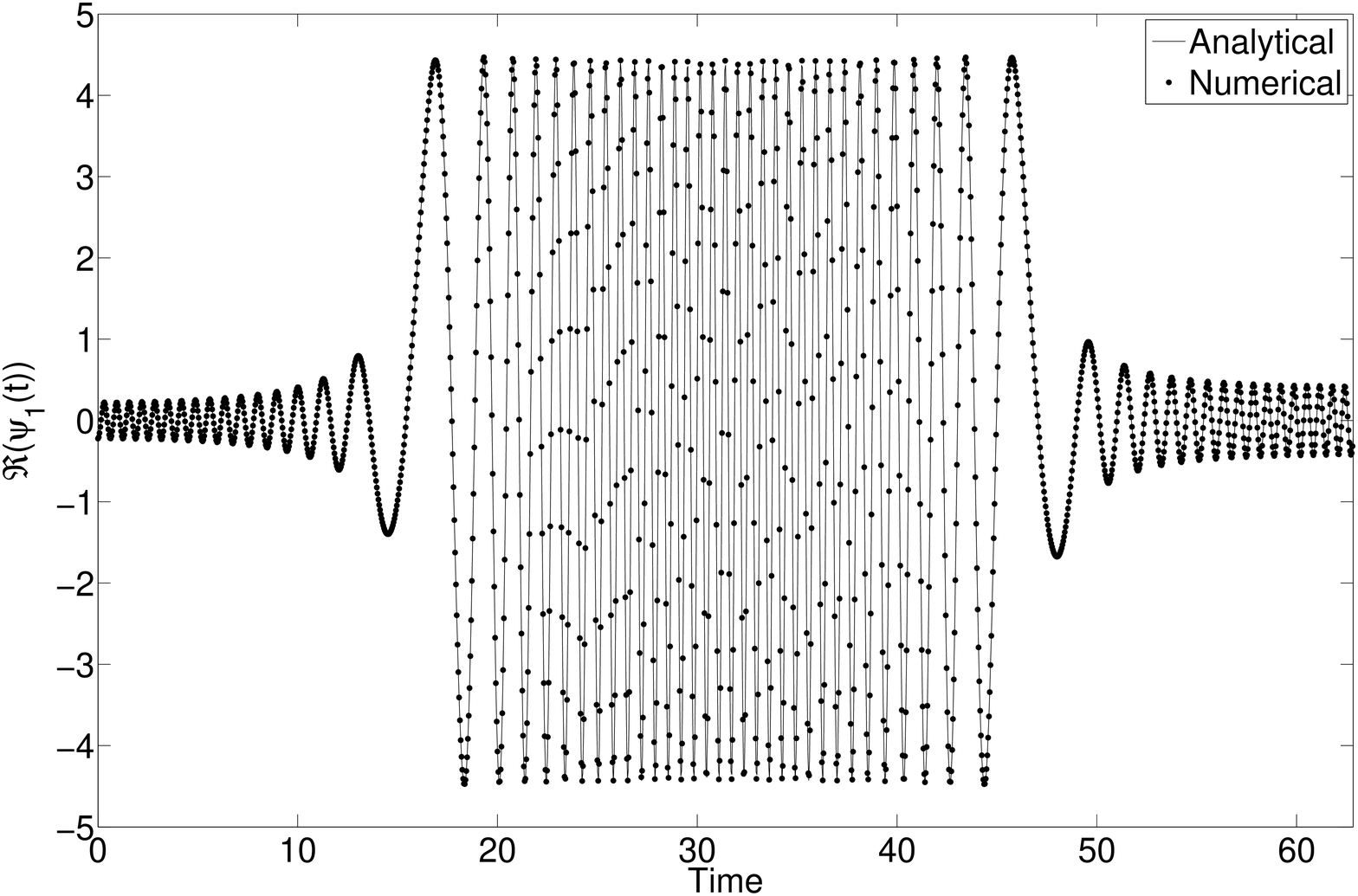}
\caption{Real part of the first component of the wave function as a function of time, over one cycle of the electric field. The analytical solution is compared to the numerical solution. The frequency is $\omega =0.1$, the momentum is $p=0$ and the field strength is $F=1.0$.}
\label{fig:WF_psi1_re}
\end{figure}

The wave function obtained from Eq. \eqref{eq:sol_H1} is plotted in Fig. \ref{fig:WF_psi1_re} for one laser cycle and is compared to an accurate numerical solution (the numerical method is described in Appendix \ref{ann::num_method}): both give the same result up to numerical errors. The constants $A^{(n)},B^{(n)}$ are determined by using the continuity of the wave function and its first derivative at points $t_{n}$.


\section{Adiabatic-impulse model}
\label{sec:adia_model}

Before computing numerical results obtained from the exact solution, it is interesting to look more closely at the analogy with the two-level quantum system to gain a better understanding of the pair creation process and interference effects. In the adiabatic regime, where the laser frequency or photon energy $\hbar \omega$ is much less than the gap $2mc^{2}$ and/or the magnitude of the vector potential, i.e. when \cite{PhysRevA.55.4418} 
\begin{eqnarray}
\label{eq:adia_cond}
4m^{2}c^{4} +\frac{F^{2}}{\omega^{2}} \gg \hbar^{2} \omega^{2},
\end{eqnarray} 
it is possible to obtain an accurate approximation of the wave function using the well-known adiabatic-impulse model described in \cite{Shevchenko20101}. The main advantage of this approach is that it yields very simple formula for the transition probabilities such as Landau-Zener formula, that allow to obtain new insights into interference phenomena. This model is also relevant in our case because the conditions in Eq. \eqref{eq:adia_cond} are fulfilled in most prospected laser infrastructures aiming at electron-positron production. In a typical laser used to probe relativistic effects and pair production, the intensity would be above $I \sim 10^{24}$ W/cm$^2$, leading to an approximate field strength of $F_{\rm laser} \sim 2.7 \times 10^{15} $ V/m. Using these conservative values, the frequency should obey $\omega_{\rm laser} \lesssim 3.5 \times 10^{18}$ Hz to fulfill the conditions in Eq. \eqref{eq:adia_cond}. This implies that in high intensity lasers (with $I \gtrsim 10^{24}$ W/cm$^2$), the production of electron-positron pairs can be approximated accurately by the adiabatic-impulse model unless the laser frequency is in the $\gamma$-ray frequencies.

\begin{figure}
\includegraphics[width=0.7\textwidth]{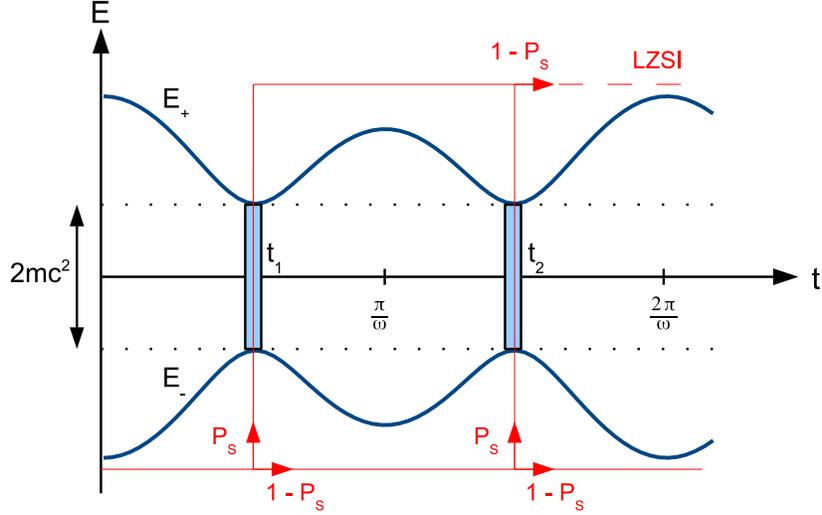}
\caption{Adiabatic energies in the driven two-level model. The nonadiabatic transitions occurs at times $t_{1,2}$. The probability of transition is given by $P_{S}$ while the probability of staying in the same state is $1-P_{S}$. In red are the different paths for transitions from negative to positive energy states. After one transition (at $t_{2}$ and all times afterwards), the part of the wave function in the negative energy states that transits upward, interferes with the part of the wave function in the positive energy states. This is the LZSI.}
\label{fig:adiabatic}
\end{figure}

Our description of the adiabatic-impulse model starts by considering the adiabatic energy of the system given by
\begin{eqnarray}
\label{eq:adia_levels}
E_{\rm adia.}^{\pm}(t) = \pm \sqrt{\left( cp + \frac{F}{\omega} \cos (\omega t) \right)^{2} + m^{2}c^{4}} . 
\end{eqnarray}
The adiabatic energies are plotted in Fig. \ref{fig:adiabatic}. In this approach, it is assumed that the wave function evolves adiabatically at all times except at points where the energy difference $\delta E_{\rm adia. }(t) \equiv E_{\rm adia.}^{+}(t) - E_{\rm adia.}^{-}(t)$ is minimal; at these points, the system undergoes non-adiabatic transitions \cite{Shevchenko20101} and the energy difference is $\delta E_{\rm adia. }(t_{1,2})= 2mc^{2}$. These points correspond to the position of avoided crossings and are given by $\omega t_{1} = \arccos \left( -\frac{cp \omega}{F} \right)$ and $\omega t_{2} = 2\pi - \omega t_{1}$ (see Fig. \ref{fig:adiabatic}). It should be noted here that these conditions are realized only if $|p|<F/c \omega$, implying that non-adiabatic transitions do not occur for large momenta since then, the adiabatic levels \eqref{eq:adia_levels} are well separated at all times.

In the adiabatic-impulse model, the time evolution is split in two parts (with $\epsilon$ a small positive time):
\begin{enumerate}
	\item Adiabatic evolution: for $t\in [0,t_{1}-\epsilon]$, $t\in [t_{1}+\epsilon,t_{2}-\epsilon]$ or $t\in [t_{2}+\epsilon,T]$.
	
	Then, the wave function is given by Eq. \eqref{eq:wf_adia} which is the adiabatic wave function.
	
	\item Non-adiabatic evolution: for $t\in [t_{1}-\epsilon,t_{1}+\epsilon]$ and $t\in [t_{2}-\epsilon,t_{2}+\epsilon]$: 
	
	Then, the wave Dirac equation can be linearized close to $t_{1,2}$. The resulting equation has a solution in terms of parabolic cylinder function (see Appendix \ref{app:non_adia})
\end{enumerate}
It is then possible to match the wave functions in the two regimes by looking at the asymptotic behavior of each solution at a time $t_{a}$ (which obey $t_{LZ} \ll t_{a} \ll |t_{1}-t_{2}|$) and solving for the integration constants (see Appendix \ref{app:non_adia} and \cite{Shevchenko20101} for calculation details, and ref. \cite{doi:10.1080/00268977000101041} in the molecular physics context). The result of this procedure can be casted in a very compact notation using transfer matrices. The final result is that (for $T \in [t_{2},t_{1}+2\pi/\omega]$ ):
\begin{eqnarray}
\mathbf{B}(T) = U(T,t_{2})N U(t_{2},t_{1})N U(t_{1},0) \mathbf{B}(0)
\end{eqnarray}
where the vector 
\begin{eqnarray}
 \mathbf{B}(t) = 
 \begin{bmatrix}
 B^{+}(t) \\
 B^{-}(t)
 \end{bmatrix} ,
\end{eqnarray}
contains the integration constants of the adiabatic solution. Note that in our case, given the initial condition of the wave function in Eq. \eqref{eq:wf_total}, we have that $\mathbf{B}(0) = (0,\sqrt{2E(0)})^{\rm T}$. The adiabatic time evolution is generated by the operator
\begin{eqnarray}
U(t_{f},t_{i}) \equiv \exp \left[ -i \sigma_{z} \int_{t_{i}}^{t_{f}} E^{+}_{\rm adia }(t)dt \right],
\end{eqnarray}
while the non-adiabatic evolution is characterized by the matrix $N$ defined in Eq. \eqref{eq:N}.

Using this result, it is straightforward to compute the rate $d\langle n \rangle/dp$. We obtain:
\begin{itemize}
	\item For $T \in [0,t_{1}]$:
	\begin{eqnarray}
	\label{eq:adia_1}
	\frac{d\langle n \rangle}{dp} = 0.
	\end{eqnarray}
	Initially, the system only has negative energy states and no transition to the positive energy states occurs adiabatically.
	\item For $T \in [t_{1},t_{2}]$:
	\begin{eqnarray}
	\label{eq:adia_2}
	\frac{d\langle n \rangle}{dp} = \frac{1}{2\pi} \frac{E^{\rm out}_{p}}{E_{p}^{\rm in}} P_{S}
	\end{eqnarray}
	When the time reaches $t=t_{1}$, there is an non-adiabatic transition from the negative to the positive energy states with a probability $P_{S}$.
	\item For $T \in [t_{2},t_{1}+ 2\pi / \omega]$:
	\begin{eqnarray}
	\label{eq:adia_3}
	\frac{d\langle n \rangle}{dp} = \frac{1}{2\pi} \frac{E^{\rm out}_{p}}{E_{p}^{\rm in}}4P_{S}(1-P_{S}) \cos^{2}(\chi + \tilde{\phi})
	\end{eqnarray}
	where $\chi = \int_{t_{1}}^{t_{2}} E^{+}(t)dt$ and $\tilde{\phi}$ is Stoke's phase defined in Eq. \eqref{eq:stokes_ph}. 
	When the time reaches $t=t_{2}$, there is another non-adiabatic transition. The wave function coming from the negative energy states interfere with the part of the wave function already present in the positive energy state, creating an interference pattern characterized by $\cos^{2}(\chi + \tilde{\phi})$. This is the essence of LZSI and is depicted in Fig. \ref{fig:adiabatic}.  
\end{itemize}
These features will be seen explicitly in the next section where the pair production will be evaluated numerically. 

Note also that these formula can be used to evaluate $\langle n \rangle$ at later times by applying the matrices $U$ and $N$. For instance, for $j$ laser cycles when $T \in [t_{2}+2j\pi/\omega,t_{1}+2(j+1)\pi/\omega]$, this would be given by
\begin{eqnarray}
\mathbf{B}(T) = U(T,t_{2})N\left[U(t_{2},t_{1})N U(t_{1},t_{2})N\right]^{j} U(t_{2},t_{1})N U(t_{1},0) \mathbf{B}(0),
\end{eqnarray}
for $j \in \mathbb{Z}^{+}$. An explicit expression of the matrix $\left[U(t_{2},t_{1})N U(t_{1},t_{2})N\right]^{j}$ is given in \cite{Shevchenko20101}.

%
%
%
%

\section{Numerical results}
\label{sec:num_res}

\begin{figure}
\includegraphics[width=0.8\textwidth]{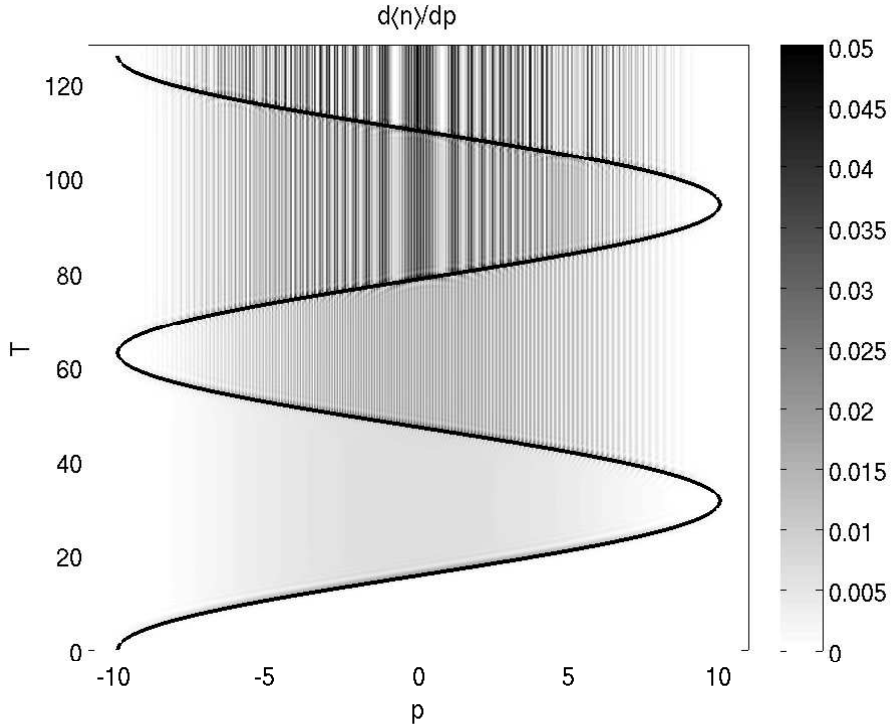}
\caption{Rate of pairs produced from the counterpropagating laser. The frequency is $\omega =0.1$ and the field strength is $F=1.0$, which insures that the system is in the adiabatic regime. The black line shows the position in $(p,T)$-space where the non-adiabatic transitions take place.}
\label{fig:pair_prod}
\end{figure}

\begin{figure}
\includegraphics[width=0.7\textwidth]{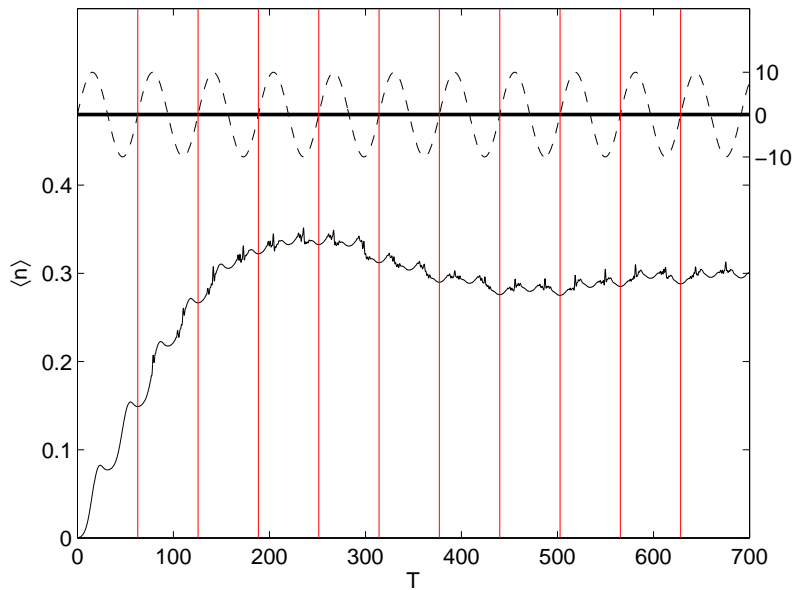}
\caption{Average number of pairs produced and electric field $E(t)$. The frequency is $\omega =0.1$ and the field strength is $F=1.0$.}
\label{fig:npair01}
\end{figure}

In this section, the rate of electron-positron pair production is calculated numerically using the exact solution and the adiabatic-impulse model. The first result concerns the quantity $d\langle n \rangle / dp$ in the adiabatic regime, which is plotted in Fig. \ref{fig:pair_prod}. In this figure, we also include the position where non-adiabatic transitions take place. It is clear from this figure that qualitative changes occur at these points. This can be understood very clearly by looking at the theoretical results obtained from the adiabatic-impulse model, in Eqs. \eqref{eq:adia_1} to \eqref{eq:adia_3}. In the first instants, there is no pair production because the system starts in a negative energy states and there is no transition to the positive energy states when the wave function evolves adiabatically. For a given momentum $p$, when the time reaches $t=t_{1}$, there is a non-adiabatic transition and pairs start to be produced with a rate given approximately by Eq. \eqref{eq:adia_2}. Later in the time evolution at $t=t_{2}$, a second transition happens and interferes with the preceding one, resulting in an interference pattern described by Eq. \eqref{eq:adia_3}. This is the well-known LZSI. After this, each time the system crosses a non-adiabatic transition, a part of the negative energy states traverses to the positive ones and a different interference pattern emerges. The corresponding average number of pairs produced (the spectrum is integrated on $p$ at each time) in shown in Fig. \ref{fig:npair01}.

\begin{figure}
\includegraphics[width=0.9\textwidth]{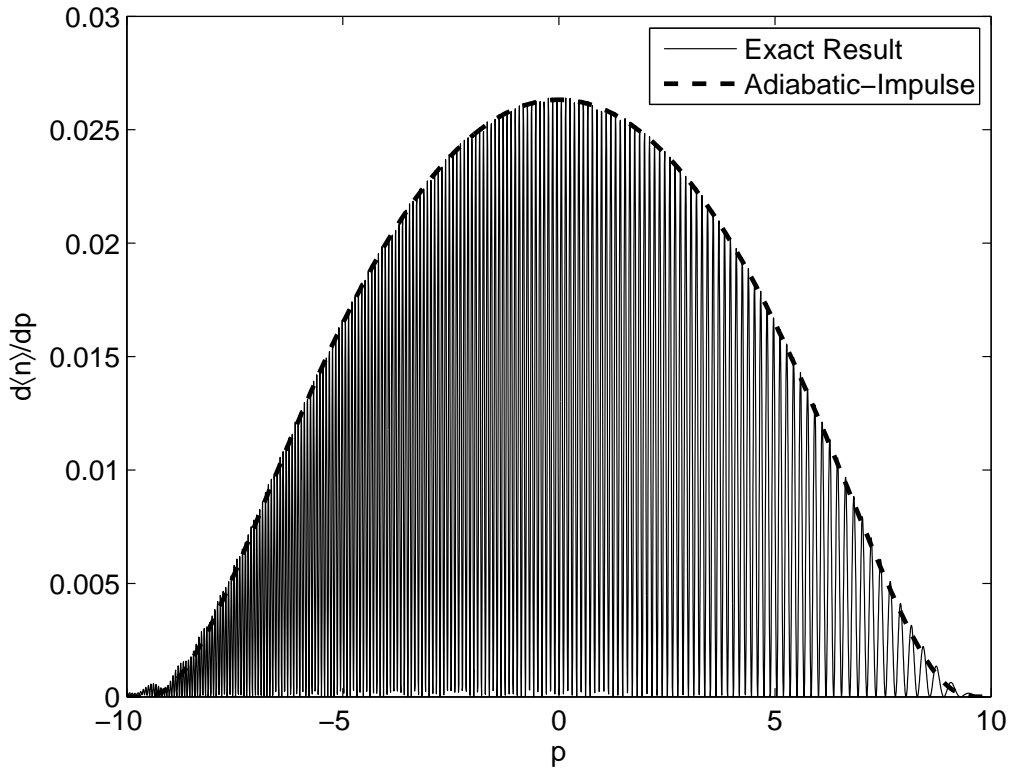}
\caption{Spectrum of pairs produced at $T=2\pi/\omega$. The frequency is $\omega =0.1$ and the field strength is $F=1.0$. The exact result is compared to the envelope obtained from the adiabatic-impulse approximation.}
\label{fig:spectrum_T1cyc}
\end{figure}

These results show a very good qualitative agreement between the two theoretical approaches. To investigate this comparison more quantitatively, the pair production spectrum  after one laser cycle (for $T=2\pi/\omega$) is plotted in Fig. \ref{fig:spectrum_T1cyc}, along with the prediction of the adiabatic-impulse model (more precisely, the envelope obtained from Eq. \eqref{eq:adia_3}). The spectrum shows the characteristic peak-valley structure of an interference pattern, in agreement with the results obtained in \cite{PhysRevD.83.065028,PhysRevLett.108.030401,PhysRevLett.104.250402}. The latter is well described by Eq. \eqref{eq:adia_3} (it was also verified that the maxima and minima of the spectrum corresponds to those of Eq. \eqref{eq:adia_3}) and thus, this effect is due to the LZSI. Also, it should be noted that again, both theoretical approaches yields very similar results.

\begin{figure}
\includegraphics[width=0.8\textwidth]{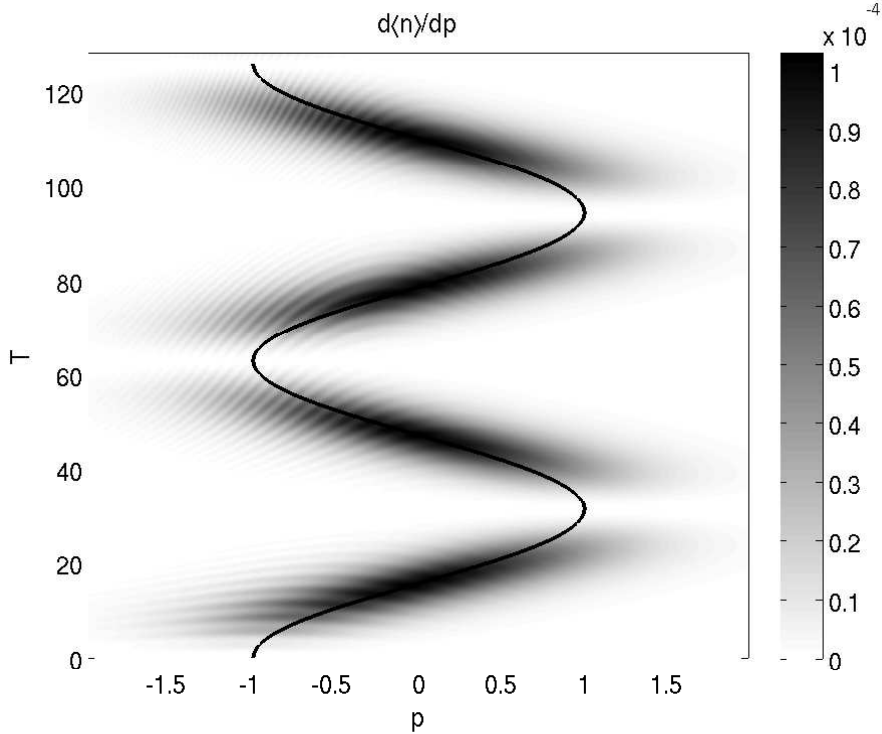}
\caption{Rate of pairs produced from the counterpropagating laser. The frequency is $\omega =0.1$ and the field strength is $F=0.1$. The qualitative behavior is different from the adiabatic regime. The black line shows the position in $(p,T)$-space where the non-adiabatic transitions take place.}
\label{fig:pair_prod_T01}
\end{figure}

Finally, in Fig. \ref{fig:pair_prod_T01}, the rate of pairs produced is presented for a smaller value of field strength ($F=0.1$). In this case the magnitude of the interference patterns are decreased significantly and the spectrum is peaked at the value of the non-adiabatic crossings. The most likely explanation for this behavior is related to the non-adiabatic transition time, which can be estimated as \cite{PhysRevLett.62.2543,PhysRevA.55.4418,Shevchenko20101}
\begin{eqnarray}
t_{\rm n.-a.} \sim \frac{mc^{2}}{F}.
\end{eqnarray}
The adiabatic-impulse model requires that $t_{\rm n.-a.} \ll \pi/\omega$ (the transition time should be much shorter than a half-cycle) to make sure that each transition is independent and well separated from each other in time. Clearly, for $F=\omega=0.1$, this condition is not fullfilled and the impulse-model is not valid for these parameter values.  


\section{Conclusion}
\label{sec:conclu}

In this article, the production rate of electron-positron pairs from high intensity linearly polarized counterpropagating lasers has been considered. An exact solution of the Dirac equation in terms of Heun's function and numerical methods has been used to compute this observable. The results has been compared to the ones obtained from another theoretical approach called the adiabatic-impulse model. The latter is based on the adiabatic approximation and allows to obtain simple expressions for the wave function. The results obtained from both methods were very consistent with each other in the adiabatic regime and when the non-adiabatic transition time is much shorter than a half-cycle. Therefore, it has been concluded that the adiabatic-impulse model is an accurate theoretical tool that may be used for other more complex systems. It was demonstrated that pair production occurs through periodic non-adiabatic transitions and that these transitions resulted in a complex interference pattern in the pair spectrum. This is very similar to the results obtained in \cite{PhysRevD.83.065028,PhysRevLett.108.030401,PhysRevLett.104.250402}. This phenomenon has been related to the well-known LZSI by using the formal analogy of our Dirac equation with the DTLS.

\appendix

\section{Numerical method}
\label{ann::num_method}

The numerical method used in this work is inspired from the spectral methods developed in \cite{PhysRevA.59.604,Mocken2004558,Bauke2011,Mocken2008868}, where an operator splitting scheme is used in momentum space to evolve the solution in time. Other approaches where the Dirac equation is solved in momentum space can be found in \cite{PhysRevA.53.1605,0953-4075-30-13-001}. The solution of the Dirac equation in Eq. \eqref{eq:dirac_eq_pot} can be written formally as
\begin{eqnarray}
 {\Psi}(t,p) &=& T \exp \biggl\{ -i \int_{0}^{t} dt' \biggl[ \sigma_{z}\left(cp + \frac{F}{\omega} \cos(\omega t') \right) + \sigma_{x} mc^{2}  \biggr] \biggr\} {\Psi}(0,p)
\end{eqnarray}
where ``$T$'' stands for the time-ordering operator. This expression can be used to solve the Dirac equation numerically in momentum space. First, the time domain is separated into $N$ time increment having a size $\delta t$. Then, it can be shown that the last expression can be approximated by
\begin{eqnarray}
 {\Psi}(t,p) &=& W(t,t_{N})W(t_{N},t_{N-1}) \cdots W(t_{1},t_{0}) {\Psi}(0,p) + O((\delta t)^{3})
\end{eqnarray}
where $t_{j} \equiv j \delta t$ and the evolution operators are given by
\begin{eqnarray}
 W(t_{i},t_{i-1}) &=& \exp \biggl\{ -i \biggl[  (\sigma_{z}cp + \sigma_{x} mc^{2})\delta t + \sigma_{z}\int_{t_{i-1}}^{t_{i}} dt'  \frac{F}{\omega} \cos(\omega t')   \biggr] \biggr\}.
\end{eqnarray}
The last equation can be computed explicitly by using the properties of Pauli matrices. It can then be written as:
\begin{eqnarray}
 W(t_{i},t_{i-1}) &=& \mathbb{I}_{2} \cos (a) -i \frac{a_{x}\sigma_{x} + a_{z}\sigma_{z}}{a} \sin(a)
\end{eqnarray}
where
\begin{eqnarray}
 a_x &=& mc^{2}\delta t, \\
a_z &=& cp\delta t + \frac{F}{\omega^{2}} \left[ \sin(\omega t_{i}) - \sin(\omega t_{i-1}) \right],  \\
a &=& \sqrt{a_{x}^{2}+a_{z}^{2}}.
\end{eqnarray}
This results in a numerical method for which the error is $O(\delta t^{3})$ \cite{pechukas:3897}.

\section{Solution at the non-adiabatic transition}
\label{app:non_adia}

In this appendix, the transfer matrix around a non adiabatic transition is derived, following the work exposed in \cite{child2010molecular,doi:10.1080/00268977000101041,nakamura:4031,Shevchenko20101} (and references therein).

In the neighborhood of $t_{1,2}$, the vector potential can be linearized in $t'$ and the resulting Dirac equation is given by
\begin{equation}
\label{eq:lin_dirac}
 i\partial_{t'}{\psi}(t',p) = \left[\mp c \sigma_{z} vt'  + \sigma_{x} mc^{2}   \right] {\psi}(t',p), 
\end{equation}
for $t_{1,2}$, respectively, and where $t'=t-t_{1,2}$ (in the following, we suppress the prime notation). Here, we have $v \equiv F\sqrt{1- \frac{c^{2}\omega^{2}p^{2}}{F^{2}}}$. The last equation is formally equivalent to the Landau-Zener transition which has a well-known solution in terms of parabolic cylinder function. The latter can be found by writing the last equation componentwise and by decoupling the two resulting equations. This yields
\begin{eqnarray}
\label{eq:2nd_order}
 \left[ \frac{d^{2}}{dt^{2}} + v^{2}t^{2} -iv+m^{2}c^{4} \right] \psi_{1}(t) = 0, \\
\label{eq:psi2_cyl}
\psi_{2}(t) = \frac{1}{mc^{2}} \left[ i\frac{d}{dt} + vt \right] \psi_{1}(t).
\end{eqnarray}
The solution of Eq. \eqref{eq:2nd_order} is found by a change of variable given by $z = \sqrt{2v}e^{i \frac{\pi}{4}} t$, which transforms the equation to
\begin{eqnarray}
 \left[ \frac{d^{2}}{dz^{2}} - \frac{z^{2}}{4} - \frac{1}{2} - i \delta \right] \psi_{1}(z) = 0,
\end{eqnarray}
where we defined $\delta \equiv \frac{m^{2}c^{4}}{2v}$. The last equation has a solution given by \cite{DLMF}
\begin{eqnarray}
 \psi_{1}(t) = C_{1} D_{-1-i\delta} \left(\sqrt{2v}e^{i \frac{\pi}{4}} t  \right) + C_{2} D_{-1-i\delta} \left(\sqrt{2v}e^{-i \frac{3\pi}{4}} t  \right),
\end{eqnarray}
where $C_{1,2}$ are integration constants and $D_{\nu}(z)$ is the Whittaker parabolic cylinder function. It is then a straightforward calculation to obtain the second component using Eq. \eqref{eq:psi2_cyl} and the recurrence relation of the parabolic cylinder functions. We get that
\begin{eqnarray}
 \psi_{2}(t) = -\frac{C_{1}}{\sqrt{\delta}}e^{-i\frac{\pi}{4}} D_{-i\delta} \left(\sqrt{2v}e^{i \frac{\pi}{4}} t  \right) + \frac{C_{2}}{\sqrt{\delta}}e^{-i\frac{\pi}{4}} D_{-i\delta} \left(\sqrt{2v}e^{-i \frac{3\pi}{4}} t  \right) .
\end{eqnarray}
The next step is the evaluation of the wave function far from the non-adiabatic transition region, that is when $t = |t_{a}| \gg 1$. Using the asymptotic expansions for $D_{\nu}$, we obtain
\begin{eqnarray}
\label{eq:asymp_sol_1}
 \lim_{t\rightarrow t_{a}} \psi_{1}(t) & \sim & C_{2} \frac{\sqrt{2\pi}}{\Gamma(1+i\delta)} e^{-\frac{\pi}{4}\delta} e^{i\phi(t)} ,\\
\label{eq:asymp_sol_2}
 \lim_{t\rightarrow t_{a}} \psi_{2}(t) & \sim & \left[ -C_{1} e^{\frac{\pi}{2}\delta} + C_{2} e^{-\frac{\pi}{2}\delta} \right] \frac{e^{-i\frac{\pi}{4}-\frac{\pi}{4}\delta}}{\sqrt{\delta}} e^{-i\phi(t)} ,\\
 \label{eq:asymp_sol_3}
 \lim_{t\rightarrow -t_{a}} \psi_{1}(t) & \sim & C_{1} \frac{\sqrt{2\pi}}{\Gamma(1+i\delta)} e^{-\frac{\pi}{4}\delta} e^{i\phi(t)} ,\\
 \label{eq:asymp_sol_4}
 \lim_{t\rightarrow -t_{a}} \psi_{2}(t) & \sim & \left[ -C_{1} e^{-\frac{\pi}{2}\delta} + C_{2} e^{\frac{\pi}{2}\delta} \right] \frac{e^{-i\frac{\pi}{4}-\frac{\pi}{4}\delta}}{\sqrt{\delta}} e^{-i\phi(t)},
\end{eqnarray}
where we defined the time-dependent phase as
\begin{eqnarray}
 \phi(t) \equiv \frac{vt^{2}}{2} + \delta \ln \left( \sqrt{2v}t \right).
\end{eqnarray}
We would like to match these asymptotic wave functions to the adiabatic wave function far from the transition times $t_{1,2}$. The adiabatic wave function is obtained as follow.

In the adiabatic approximation, the wave function looks like
\begin{eqnarray}
 \psi_{\rm adia.}^{\pm}(t) = \varphi^{\pm} \exp \left[ \mp i \int^{t}E(t') dt' \right],
\end{eqnarray}
where the $\pm$ denotes positive and negative energy solution, respectively, and $\varphi^{\pm}$ are adiabatic coefficients to be determined which obey $|\partial_{t} \varphi^{\pm}| \ll |E(t) \varphi^{\pm}|$ for all times. By substituting this into the Dirac equation
\begin{eqnarray}
 i\partial_{t} \psi(t) = \left[ \sigma_{z} P(t) + \sigma_{x}mc^{2} \right] \psi(t),
\end{eqnarray}
where $P(t)$ is the canonical momentum for the system under consideration, normalizing the wave function such that $|\psi_{\rm adia.}^{\pm}|^{2}=1$, we arrive at the following general solution (which is a linear combination of positive and negative energy solutions):
\begin{eqnarray}
\label{eq:wf_adia}
 \psi_{\rm adia.}(t) = \sum_{\pm} B^{\pm} \varphi^{\pm} \exp \left[ \mp i \left(\int_{0}^{t}E(t') dt'  + \frac{\pi}{4}\right)\right],
\end{eqnarray}
where
\begin{eqnarray}
 \varphi^{+} = 
\begin{bmatrix}
 \sqrt{\frac{E+P(t)}{2E}} \\
\sqrt{\frac{E-P(t)}{2E}}
\end{bmatrix}
\;\; ; \;\;
 \varphi^{-} = 
\begin{bmatrix}
\sqrt{\frac{E-P(t)}{2E}} \\
 -\sqrt{\frac{E+P(t)}{2E}} 
\end{bmatrix},
\end{eqnarray}
and where $B^{\pm}$ are integration constants. Note that the factor $\pi/4$ in the phase appears when the next order in the adiabatic approximation is considered \cite{child2010molecular}.

Here, we are considering two times $\pm t_{a}$ which are lying in a region close to the transition region but which are much longer than the typical non adiabatic transition time $t_{\rm LZ}$, that is $t_{\rm LZ} \ll |t_{a}| \ll |t_{1}-t_{2}|$. In this region, the time evolution of the wave function can be described accurately by the linearized Dirac equation, Eq. \eqref{eq:lin_dirac} (thus, $P(t) = -vt$), while still being far from the transition times. In this case, assuming that $v|t| \gg mc^{2}$ also holds, the adiabatic solution can be simplified to give
\begin{eqnarray}
\label{eq:psi_adia_exp}
 \psi_{\rm adia.}(\pm t_{a}) \sim 
\begin{bmatrix}
 B^{\mp} e^{  i \phi(t_{a}) - i\frac{\delta}{2} \left[ \ln(\delta) - 1 \right] - i \frac{\pi}{4}}\\
\pm B^{\pm} e^{ - i \phi(t_{a}) + i\frac{\delta}{2} \left[ \ln(\delta) - 1 \right] + i \frac{\pi}{4}}
\end{bmatrix},
\end{eqnarray}
The time dependence of the last expression is the same as the asymptotic solutions in Eqs. \eqref{eq:asymp_sol_1} to \eqref{eq:asymp_sol_4}, allowing to match the solutions at $t=\mp t_{a}$ and thus, to determine the transfer matrix that allows to link the solution at negative times to the one at positive time. Solving for the integration constants $B^{\pm}$ in Eq. \eqref{eq:psi_adia_exp}, we find that the non-adiabatic transition can be characterized by the following time-independent transfer matrix \cite{nakamura:4031,Shevchenko20101}:
\begin{eqnarray}
\label{eq:N}
 N \equiv 
\begin{bmatrix}
 \sqrt{1-P_{S}}e^{-i \tilde{\phi} } & -\sqrt{P_{S}} \\ 
 \sqrt{P_{S}} & \sqrt{1-P_{S}}e^{i \tilde{\phi} }
\end{bmatrix},
\end{eqnarray}
where the Stoke's phase is defined as
\begin{eqnarray}
\label{eq:stokes_ph}
 \tilde{\phi} \equiv - \frac{\pi}{4} + \delta [\ln (\delta) -1] +  \arg \Gamma(1-i\delta).
\end{eqnarray}
The quantity $P_{S} \equiv e^{-2\pi \delta}$ is Schwinger's result for the pair probability creation in a constant field (note that it is also the Landau-Zener transition probability). The matrix $N$ connects the wave function before and after the non-adiabatic transition.

\begin{acknowledgments}
The authors would like to thank Sczcepan Chelkowski for many interesting discussions and for critically reviewing the manuscript. 
\end{acknowledgments}

\bibliographystyle{apsrev}

\bibliography{bibliography}

\end{document}